# Enhancing Tourism Destination Accessibility in Developing Countries through Virtual Worlds

**Abstract**

The problem of destination accessibility is a vital concern in the sustainable tourism development in the emerging regions due to the increasing numbers of tourism business growth in the recent times. Tourism is one of the potential foreign exchange earning sectors, which place sustainability as one of the main success metrics for benchmarking the industry's overall development. On the other hand, there are several destinations, which are inaccessible to tourists due to several reasons. Underutilization of potential destinations in both pre purchase and consumption stages is a strategic disadvantage for emerging countries on leading their tourism industry towards sustainability. This research embarks on a content analysis of second life based tourism groups and places. Requirement of a virtual world model to increase the destination accessibility of tourism products has been outlined. The model has to be designed with visual and auditory experience to tourists. The model is expected to enhance the accessibility of destinations for users of different categories.

1. ## Introduction

The question of sustainability in tourism development is a contemporary concern. In other words, developing tourism in a manner which does not compromise the preservation of tourism resources for the future generations is a critical concern. The present global credit crunch has further increased the degree of challenge involved in ensuring sustainable tourism development. The emerging countries, being vulnerable to the above dilemma need to adapt innovative methods to ensure sustainable development of their tourism business. Arguably, tourism is one of the potential foreign exchange earning sectors, which place sustainability as one of the main success metrics for benchmarking the industry's overall development. On the other hand, there are several destinations, which are inaccessible to tourists due to several reasons. Underutilization of potential destinations in both pre purchase and consumption stages is a strategic disadvantage for emerging countries on leading their tourism industry towards sustainability.

Firstly, the quality of product information largely relies on the degree of experience added with the information. In particular, the travel experience delivered online in terms of visual and auditory forms play a vital role in influencing the purchase decision of tourists. This entails the need for bridging the gap between product information and the exact experience. Improved quality of information in the pre purchase stage will increase the revenue by positively influencing the purchase behavior of consumers. Also, it will enhance the post consumption experience by reducing the mismatch between expectations and exact features. Secondly, tourists with impairments and elderly tourists get deprived from consuming the tourism product, as same as the normal tourists. This deems a need for bringing the destinations to their doorstep, with all its real world experiences to enhance tourism development in the consumption stage. Thirdly, there are inherently endangered tourism destinations, which cannot be accessed by human beings due to complex geographical location. Enablement of tourists to get access to these destinations is another open question, which could result in the development of new tourism products. Hence, a solution to enhance the accessibility of tourism destinations in all these three means is an evolving contemporary need for successful tourism development. This paper, proposes a virtual world model to bridge the above gap, along with a number of distinct value additions to tourism development. This means the outlined model could enhance the accessibility of tourism destinations to people with impairments through delivering a tourism experience in the virtual world. This will be achieved through making these countries accessible in terms of destination for the impaired person as well as ensuring the interface is accessible to all users.

## 2. Avatar Based Interactions

Interactions in Virtual Worlds are essentially about avatar interactions. Avatar interactions are quite different and unique compared to the Web 2.0 based interactions which are solely between the interface and the user. According to Davis *et al.* (2005) "An avatar is defined as a user created digital representation that symbolizes the user's presence in a metaverse". Especially the need for Avatar comes in to play when the user needs a digital representative for them in the virtual world. Avatars are used as a form of interaction agents in the virtual worlds which could be customized according to individual needs and personality types. According to Kohler *et al.* (2009) avatars are the vehicle of the self and are inhabitants of virtual worlds. The second life is used as an innovation platform where avatars could be used to develop products in the second life platform.

Some companies are using avatar based innovations as a way forward to new product development. To clarify, the avatar-based approach adapted in the above context opens up avenues for innovative product design and content creation in the virtual world (Kohler, *et al.*, 2009). They argue that the avatar based innovations present an opportunity for companies to engage with customers in new and innovative ways during an interactive new product development process. The user generated nature of the virtual platform could further enrich innovation efforts. Research on how the companies could attract appropriate customers, and which incentives the firms need to implement in order to promote and leverage valuable customer contributions, could provide some open avenues for further exploration.

Additionally, to which extent an avatar is engaging in deceptive behavior or what measures the companies could take to guarantee the quality in contributions - are some potential areas for further research. Furthermore, the nature of Virtual Worlds calls for a re-examination of various issues such as avatar motivation to engage in co-innovation activities and the degree of interaction efficiency among avatars. The question arises as to what are the mechanisms for supporting and facilitating collaborative innovation in Virtual Worlds? According to Kohler *et al.* (2009), further research is also warranted to compare traditional web based methods with avatar based efforts, to shed light on the question of when to best employ which technology. It is also important to study their use for diverse new product development tasks due to the market forces, niche opportunities, technological development, and so on. However, this research does not appear to focus on the interaction related issues pertaining to the avatar based innovations. Particularly, the psychological concerns pertaining to the avatar based interactions will pose challenges to the effectiveness of interactions compared to real world interactions.

Another important use of Virtual Worlds is the ability to collaborate in real time through synchronous communication medium. Virtual Worlds allow globally dispersed teams to collaborate and work in a virtual environment, where the avatar based interaction happens as part of the collaboration (Davis et al., 2009).

A recent study explores on the potential of Virtual Worlds in enriching innovation and collaboration in Information Systems research, development and commercialization (Dreher et al., 2011). The authors argue that the Virtual Worlds - by their very structure provide a powerful context for innovation and collaboration. Their paper concludes stating that there is great potential inherent in the use of 3D Digital Ecosystems for Information Systems Technology research, development, and commercialization. Such developments will keep pace with the digital-native culture of younger generations and have the potential to innovatively revolutionize our social systems relating to governance, education, commerce, and social interaction. Digital Ecosystems, 3D Virtual Worlds in particular, are set to lead the charge in our modern culture of accelerating innovation (Dreher et al., 2011).

Another recent study by Eklund *et al.* (2009) reports, a Virtual Museum of the Pacific - implemented as Web 2.0 application that experiments with information and knowledge acquisition for a digital collection of museum artifacts from the Australian Museum. Hence the mission of this paper is to evaluate the emerging Virtual World models with regard to the socio-political, technological and ethical aspects and to evaluate the degree of contributions made by these emerging models to the body of knowledge.

Messinger *et al*. (2008) discussed the typology for Virtual Communities, and the historical developments of Virtual Worlds research, clearly outlining the development of the gaming industry as well as the social networking industry. The field of Virtual Worlds is a unique blending of both these industries over a few decades. The paper further discusses the 5 Ps of a Virtual Community namely purpose, place, platform, population and profit model. Finally the paper uses the typology to interpret the development in the social networking and electronic gaming industries. The paper further claims the above typology will be useful in figuring out the upcoming trends for other industries as well. This typology has been utilized for evaluating the virtual world cases from the literature during the course of this research.

## 2.1. 3 Dimensional Virtual Ecosystems

A study conducted at Curtin Business School evaluated using second life as a 3 dimensional virtual ecosystem for information systems development and research commercialization (Dreher *et al., 2011*). The study focused on deploying the Automated Essay Grading (AEG) system called MarkIT™, developed at Curtin Business School, in Second Life. The paper builds an interesting case of commercializing the AEG technology through the use of 3D Virtual Worlds. In the above case, the purpose of interaction is to enhance research efforts and commercialize research through the use of second life. The place of interaction is in second life with the support of virtual world platform. The population includes the researches from both Curtin University as well as the Graz University of Technology. Finally the return on interaction goes up to increased research productivity and very high quality collaboration evident by increased research funding and effective publications.

### 2.2. Automated Assessment Laboratory

Another concurrent study conducted at Curtin Business School evaluates on the potential of second life in Information systems education. The system evolves into the concept of utilizing 3D Digital Ecosystem in second life based application - the content is generated entirely by the user. This promotes a rich culture of innovation around the application which could be further developed into capabilities for 3D Digital Ecosystem. The application has been established as 3D Digital Ecosystem. There are several benefits and few limitations involved in using 3D Digital Ecosystems in e-learning. Mainly, the geographically dispersed students studying systems development course will have an advantage of getting exposed to real time practical sessions on systems development through the second life based classroom. And the efficiency of teaching process is increased due to the interactive avenues provided by second life in teaching and learning. Also, learning systems development in 3DDE allows geographically dispersed student population to learn systems development by doing. In contrast the system needs high infrastructure facilities for implementation. This is a major challenge for developing countries as well as for the remote regions of the developed countries.

The purpose of AEG technology developed at Curtin Business School was to facilitate the assignment assessment and moderation through an automatic tool which would be further extended to manage the entire learning process from end to end. The interaction takes place in the second life space dedicated for Assignment Assessment library. The platform is Second Life and the population consists of students following the Information Systems degree program at Curtin University and Graz University of Technology. The profit model of this application was through subscriptions. An innovative business called Blue Wren was established to commercialize the AEG technology.

### 2.3. Virtual Museum of the Pacific

The virtual museum of the pacific is another interesting case to be evaluated. The authors describe the design process adapted for building the digital museum project hosted at the University of Wollongong. The paper describes the Virtual Museum of the pacific as a digital ecosystem in which objects of a digital collection of museum artifacts are derived from facets of the physical artifacts held in the Australian Museum's pacific collection. The virtual museum of the pacific allows several diverse search methods: attribute search based on a control vocabulary, search via query refinement and query by example. Further to this the system also provides a number of management interfaces that enable content to be added and tagged, the control vocabulary to be extended, user perceptions to be defined and narratives added via wiki.

In evaluating the typological aspect of this case, the purpose is defined to be as providing museum access to the large audience through the virtual museum access to large audience through the virtual museum interface. The place of interaction is set to be completely Virtual through the interface designed as part of this project. Platform of this case is through a synchronous communication mechanism facilitated through a query refinement approach. The population or pattern of interaction in this case is defined as the large target audience focuses through the museum. Finally, the project model would be the subscription fee or payments such as endowments provided by the user as part of / as a result of their interaction.

### 2.4. Virtual Therapeutic Community

Good *et al.*, (2011) explored on a virtual therapeutic model for support and treatment of people with Borderline Personality Disorder. The paper evaluates the position of an ongoing investigation into developing the requirements and acceptance of a Virtual therapeutic community hosted in second life. The propose model is evaluated according to the typological elements of a virtual community.

Firstly, the purpose of interaction is to support patients with Borderline Personality Disorder. The purpose could be achieved through by the model, in case it satisfies the information needs of the people with Borderline Personality Disorder. The second typology – place is critical for the success of the aforementioned therapeutic model. Particularly, the place of the therapeutic model could be the virtual world space hosted in second life. The interaction takes place in the second life environment - which could be further expanded to the web 2.0 based information sites to support the information needs of the people with BPD. Thirdly, the platform is considered. As per the published papers on the above model, there is no defined platform is yet there for this therapeutic model. The type of communication platform (i.e. synchronous / asynchronous) has to be finalized yet for the above platform. Fourthly, the population of the above model is predefined to include patients with BPD. Also the population may further extended up to including patients with other psychological disorders in the forthcoming stages of the project evaluation. Finally, the profit model of this second life based virtual therapeutic community has not been defined so far. However, the model is based on the Henderson hospital model – which is one of the well know therapeutic models so far with regard to treating people with Borderline Personality Disorder. To conclude, the second life based virtual therapeutic community model is still in its inception stages, but yet has the potential to be used as a powerful Virtual World application for treating people with mental illness in long term.

### 2.5. Otago Virtual Hospital

Otago Virtual Hospital model is designed to formatively assess dispositional behaviors in scenario based in the Virtual Worlds. The framework was devised for use with medical students playing the roles of junior doctors as they solve open ended clinical cases within an environment called the Otago Virtual Hospital. In doing this, the authors designed a conceptual framework in which medical students are retrospectively assessed based on the number of times they either seized or missed an opportunity to engage in a particular dispositional behavior. In this article, the authors have presented an empirical illustration of our conceptual framework. The purpose of Otago Virtual Hospital is to formatively assess dispositional behaviors in scenario based in the Virtual Worlds. The place is Second Life space dedicated for the hospital. The platform is the second life based virtual world in which all the interactions take place. The population of the above model covers medical students playing the roles of junior doctors. Finally, the profit model of Otago Virtual Hospital is through normal revenue generated through increased service satisfaction and enhanced efficiency.

### 2.6. Virtual Physics World

Wegener *et al.* (2012) reports the implementation of virtual reality software which has similar characteristics of game which could be utilized for teaching physics. The mechanism is utilized for modeling students' usage of the package through different types of methodologies adapted throughout the study. The teaching package includes several aspects which are utilized to support the key assessment methodologies in the mechanism. This model is a unique contribution to the literature on the field. However, this model only focuses on the relativity simulation – which is an integral part of theoretical physics. There needs to be similar simulation models created for explaining other core theories in physics. However, when comparing with the information system field, the model is expected to be useful for students in understanding and comprehending the most difficult conceptual elements of theoretical physics. Also, another strength associated with this model is its student – centeredness.

### 2.7. Virtual Birth Centers

Stewart *et al.*, (2012) describes a virtual birth center created on second life for the purpose of training midwives on the same. This is a New Zealand based initiative focused on improving the midwife education. The model creates a number of clinical scenarios through which the midwifery training could be delivered in immersive engagements. The study is also used as a case for evaluating the effectiveness of Multi User Virtual Environments (MUVE) based teaching and learning initiatives. On the contrary, the Virtual Birth Centre (VBC) has been criticized for not been used for several years. Another challenge identified with this project is the unlikely partnership between stakeholders (i.e. midwives and IT specialists). The successful collaboration between these parties becomes critical elements for the successful implementation of the project; since the mutual understanding of each party regarding the capabilities of others is important for successful implementation and collaboration. Also there are a number of issues identified with regard to the practices and present policies on the sustainability of VBC. Although, there has been reported collaborative work between institutions in this area, still there is a room for collaboration and joint research as per the authors. It may well be suitable to say that the collaborations in this area need to be further enhanced in order to have a successful research endeavor on the area. Also, a strategic approach is an essential requirement in this project. This is an obvious but essential point as the strategic approach will enhance the efficiency of implementation as well as in tackling the post implementation challenges which will eventually leads to a sustainable system.

## 2.8. Improved access to 3D Virtual Learning Environments

Wood and Willms (2012) investigated the potential of an accessible 3 dimensional virtual learning environment with the participation of students with disabilities. The finding shows that there has been a plethora of research on providing access to virtual environments for people with disabilities. The findings are also applied across the educational industry for increasing the accessibility of such systems for users with disabilities. On the contrary, a study by Good (2008) devised a new method for improving access to web based information for users with disabilities. The study yielded in a method proposing to re-order search results according to user rating of web content so that the accessibility is increased. The study further evolved on re-ordering search results according to learner styles, so that the accessibility of e-learning content would be regularized according to the learning preferences of learners with differing learning styles. In this present paper Wood and Willms (2012) focuses on the participation and access related issues faced by people with disabilities. Further research is required to investigate the potential of 3 dimensional digital virtual learning environments to enhance participation and access for students from diverse backgrounds.

## 2.9. Virtual Hats – A Role Playing Activity

Role playing in Virtual Worlds have a tremendous potential for allowing students to have effective learning endeavors due to the synchronous communication. Also, obviously Virtual Worlds have the potential to facilitate students engage in learning activities which are not possible in the real world. In this article, a project that involved pre-service teachers carrying out role-plays based on de Bono's *Six Thinking Hats* framework is presented. A pilot study was carried out over two years with on-campus students, who performed the role-plays both in a real-life, physical setting and within the virtual world of *Second Life*. In overall the study presented argues that real world setting could be simulated or replicated in the virtual words. The results of the pilot study suggest that students have a preference for real-life, face-to-face learning activities; however, the participants in the study were on-campus students, who, unlike those who are studying at a distance, actually have the privilege of access to this mode of learning.

## 2.10. Chinese Language Lesson in Virtual Worlds

Henderson et al. (2012) reports on a study evaluating the single collaborative language lesson using second life. The results shows that the single collaborative language lesson using second life could show statistically significant increase in student self-efficacy believes. However there was a difference in the durability of language performance believes among the students with different real life experience. The purpose of this case is to evaluate and enhance the language learning believes of students through the use of second life. The place of interaction is the second life space dedicated for this task. The platform is the second life based virtual world platform allocated for this interaction. The population is the language teaching students who are using second life based mechanism to learn and improve their language abilities. The profit model has not been indicated across the entire article. However, it could be assumed that the profit model could be through the subscription fees paid by the students.

### 2.11. Learning Law through Second Life

According to this study, the present teaching method for law education in the Australia is traditional one. Legal doctrines and policies are transmitted to students from lecturers in the traditional lecture room method at present. The programs embedded into this research facilitate a flexible learning mode for students where they could utilize these facilities to learn on their own independence. The purpose of this model is to model is to facilitate law education. The place of interaction is the second life based platform dedicated for this purpose. The platform of interaction is the second life based virtual world platform. The population consists of law students where the profit model could be through subscription fees.

### 2.12. Learning and Corporation through Virtual Worlds

Pelet *et al.*, (2011) explores on enhancing learning and corporation through digital Virtual Worlds. The paper focused on developing a second life based virtual classroom for students. The paper explains an experiment conducted on using second life as a teaching tool. It tests the feasibility of implementing Second Life in the classroom by using 168 non expert students as samples for this study. As an obvious bottleneck, non-expert students need to be trained on second life before being experimented in the second life based classroom. The study is profitable and could make potential contributions to the body of knowledge. However, there are obvious limitations observed with this study. Firstly, the experimentation was conducted for a very short time. Students were observed to be feeling frustrated in the studies due to the short period of time length used for collecting the samples.

### 2.13. Virtual Reality in Tourism Development

Steuer (1993) argues that 'presence' and 'telepresence' as the underlying conceptual elements of virtual reality. Particularly the sense of being in an environment is the main requirement, which could be generated by natural or mediated means. Cheong (1995) defines virtual reality as "a computer mediated sensory experience that serves to facilitate access into dimensions that differ from our own". However, the above definition does not have an operational level focus, but more of an abstract level statement. On the other hand, there are certain constraints involved in creating a computer mediated sensory experience. In particular, there are certain technological constraints involved in developing a system which could generate experience to satisfy all five senses. Although it is possible through the state of the art virtual reality studies, it is a challenge to implement such a system in emerging countries due to the infrastructural constraints involved. Particularly, the organizations in the emerging countries would not be able to afford huge amount of money on a virtual reality system, while there are several other priorities to be addressed while making investment decisions in tourism development. Pizam (2009) indicates that the global financial crisis has immensely impacted the global travel industry. Hence, there are certain financial constraints involved in producing a complete virtual experience of five senses through a system in the context of emerging regions. Therefore Virtual Reality is defined as "a computer mediated sensory experience which facilitate access to visual and auditory dimensions of a travel destination" in the course of developing this model. The definition provides a contextual relevance to the requirements deemed by the travel related endeavors of emerging regions.

A recent study claims game animation technology as the possible next innovation in tourism (Tjostheim et al, 2005). But the above study evaluated the context of an entire 'virtual tour' facilitated through WWW. Also this research was tested with retail business - a non travel industry. However, the user testing done by Tjostheim et al (2005, p 9) with regard to the game animation technology utilization indicates that a typical home system will be enough to facilitate the above technology. Also the study reports a high technical compatibility of the particular technology. However, contextualizing the findings to tourism domain is an obvious challenge in front of practitioners. Above findings, in light of the current period of financial downtrodden urges the Sri Lankan hoteliers to go for middle approach between a complete virtual tour and passive web presence. Consequently, a part application of simulation technology allowing a small glimpse of tours experience to be received through the website will better facilitate excellence in web tourism promotion.

| No | Case | Author | Contribution |
|---|---|---|---|
| 1 | 3D Digital Ecosystem for Information Systems Research | Dreher et al. | Creating viability for using Second Life as 3D Digital Ecosystem for Information Systems Research |
| 2 | Virtual Product Development | Kohler, et al. | Companies using avatar based innovations as a way forward to new product development. |
| 3 | Virtual Research Commercialization | Dreher et al. | Virtual Worlds - by their very structure provide a powerful context for innovation and collaboration |
| 4 | Virtual Museum of the Pacific | Eklund et al. | Implemented as Web 2.0 application that experiment with information and knowledge acquisition for a digital collection of museum artifacts from the Australian Museum |
| 5 | Automated Assessment Laboratory | Dreher et al. | To facilitate the assignment assessment and moderation through an automatic tool which would be further extended to manage the entire learning process from end to end. |
| 6 | Otago Virtual Hospital | | To formatively assess dispositional behaviors in scenario based in the Virtual Worlds |
| 7 | Virtual Physics World | Wegener et al. | To evaluate students' use of the package followed by instructional and software development |

| 8 | Virtual Birth Centers | Stewart *et al.* | To train midwives on child delivery |
| 9 | Virtual Hats - Role Playing | | To allow students to have effective learning endeavors due to the synchronous communication |
| 10 | Chinese Language Learning in VW | Henderson *et al.* | To evaluate the single collaborative language lesson using second life |
| 11 | Law Learning in Second Life | | To use of *Second Life* machinima in two interactive multimedia programs namely; Air *Gondwana* and *Entry into Valhalla* |
| 12 | Virtual Therapeutic Community | Good *et al.* | To enable patients with BPD get support in times of crisis |
| 13 | Improved access to 3D Virtual Learning Environments | Wood and Willms (2012) | Investigated the potential of an accessible 3DVLE for increasing access and participation of students with disabilities is reported, and strategies for improving outcomes |

### 3. Methodology

The methodology used for this research is of a qualitative approach with content analysis method for data analysis. The second life based virtual tourism places and groups were analyzed. Firstly, a through literature review has been conducted to evaluate the cases from the literature through the lens of 5Ps typology. Secondly, a search has been conducted on the second life based tourism places. The profile of tourism places has been recorded in a spreadsheet along with the details such as location, owner details, size, category and maturity. Secondly, the second life based tourism groups were analyzed using a search conducted for tourism groups. The details such as founder, maturity and group description were recorded in the spreadsheet. The rational for selecting these sites and groups relies on the tagline related to the tourism. The analysis process includes identifying the tourism places and groups in the second life and documenting those in order to form a clear basis for the proposed model. The new model has been proposed based on the results obtained through data analysis.

### 4. Results

**Tourism Places in Second Life**

The initial web search run through the second life search button to trace tourism related places in second life reveals that there are 13 places in the second life with the tag of tourism. The table shows the summary of the available tourism places in second life.

| Place | Owened by | Location | Size | Category | Maturity |
|---|---|---|---|---|---|
| Domun Rentals Tourism Office @ Pslande Reserved Forest | Domus - Rentals & Real Estate | Psland (96, 244, 31) | 2912 sq. m | Parks & Nature | General |
| Maasai Mara Basecamp | Massai Tourism Project | Maasai Tourism (152,110,32) | 65536 sq. m | All | General |
| France - Paris 1900 - Moulin Rouge - Champs Elysees | Admicik Grp | Paris 1900 (32, 142, 24) | 48992 sq. m | All | Moderate |
| Frame Paris - Les Champs Elysees | Admicik Grp | Les Champs Elysees (123, 11, 24) | 63008 sq. m | All | Moderate |
| Cabtel | Netrizio | Central (128, 128, 0) | 62336 sq. m | All | General |
| Moulin Rouge_Paris 1900_Troupe du Jardin de Paris | Admicik Grp | Paris 1900 (101, 179, 24) | 2080 sq. m | All | Moderate |
| Office du Tourism de Midi - Pyrenees | CRT ARDESL Metalab | Paris Ilede Frame (55, 201, 23) | 4352 sq. m | Arts & Culture | General |
| American Plaza | Library Bluebird | Library City II (130, 78, 0) | 16 sq. m | All | Moderate |
| Kansas Land | Kansas Librarians in SL | Library City II (36, 164, 22) | 4160 sq. m | Educational | Moderate |

There are less number of adult places exist in the second life. Almost all the places are either general or moderate in the maturity level of users.

**Tourism Groups in Second Life**

The initial web search run through the second life search button to trace tourism related groups in second life reveals that there are 26 groups in the second life with the tag of tourism. The table shows the summary of the available tourism groups in second life.

| Group | Founder | Maturity | Description |
|---|---|---|---|
| Sweden Tourism | Sweden Charisma | General | A group to discuss what to see when visiting Sweden, things to do, and eating / sleeping suggestions |
| Tourism Ireland | Dublin Owner | General | This group is for people involved in the tourism Ireland project in second life |
| SL Edu Teachers Tourism | Alida Juliesse | General | Teaching, Learning in Second Life |
| Aksum Immigration / Tourism | Nemesis Sass | Moderate | this group serves as the initial entry and tourist (observer group) for the Aksum Imperium |
| Discover SL Tourism Agency | DNA Alter | General | To guide SL users on exploring the SL usage |
| SCU Events & Tourism | Chancellor Shackieton | Moderate | No description available |
| AAA Tours Tourism | Dennis Mitchel | Moderate | No description available |
| Exist Tourism Project Tour Operator | Flavio Earst | Moderate | No description available |
| Massai Tourism Project | Yuchih Andel | General | No description available |
| Visit Mexico | Lester Netarious | General | No description available |
| St' Lousious Tourism - DDINC Admin | Andrew Hughes | General | No description available |
| Visit Mexico | Valiant Strangelove | General | No description available |
| Galveston Islan Cand Trust | Link Pippen | General | A friendly SIM promoting RL anlveston Island |
| West Cork Tourism Committee | Meekal Kilara | General | No description available |
| Consultant of Europa Wulfenbach | Melanippe of Themiscyra | General | To promote interest in and tourism to Europa wulfenbach, and develop friendly, mutually supportive relations |
| Spanish Speak in SL | Wilson Voight | General | Spanish classes, study groups, conversation practice partners and language tourism |
| 1920s Berlin Brothel | Mab Ashdere | Adult | Part of the 1920s Berlin project between 1921 and 1933, Berlin developed a reputation for debaucherg unralead by any city |
| USC Public Diplomacy Events | Miranda Tibbett | General | A group for events, meetings and panels focussed on public diplomacy and hosted by the USC center on public diplomacy at the annerberg school |
| Galveston Island | Link Pippen | General | This group is for managing the Galveston Island Private SIM and to promote RL Galveston island, Texas Tourism |
| Viandando | Neverstop | Moderate | Viandando is Florence (Italy) based tour operation specializes in services and events |

**Virtual World Model for Tourism Development**

Almost all the places are either general or moderate in the maturity level of users. The similar case applied to the tourism groups as well. The number of well mature (adult) groups in the second life is minimal according to the data collected through the content analysis. Therefore, there is an emerging need for a new model which comprises of the missing requirements in the already existing tourism places and groups. The model aims to enhance the tourism development of specific regions by increasing the destination accessibility, through which the economic sustainability could be dramatically improved. Specifically, the proposed model will enhance the accessibility of tourism destinations to tourists with mobility impairments as well elderly tourists. The experiential nature of VR provides more sensory and rich information about the destinations to tourists (Guttentag, 2010). Hence, the purchase decision of tourists could be immensely influenced through this model. This could lead to a major economic gain through increasing the number of visits to the destinations through virtual world. In addition to this, the model could be used as an advanced promotional tool for tourism through which a prospective tourist could gain a more advanced understanding of the features of specific tourism destinations. Also, researches argue that an e-tourism experience would immensely helps tourists with panic disorder to get rid of unnecessary dismay during the exact visit (Newman, 2008). Furthermore, the model could possibly used to substitute the travel experience of endangered destinations, which are inaccessible to even normal tourists. In conjunction, the model could also be used for environmental gains related to tourism development. Especially, the eco-efficiency of tourism destinations could be indirectly enhanced through increasing the virtual visits instead of trouping horde of people across delicate habitats. Particularly the solution could immensely contribute towards the economic development of developing countries, which earns a considerable amount of annual foreign exchange through tourism.

**Further Enhancements**

The above outlined model could be utilized for enhancing the overall development of tourism business in emerging regions. Firstly, the model could be utilized for developing virtual theme parks. Development of virtual theme parks has already been identified as a potential benefit of virtual reality in tourism business (Williams & Hobson, 1995). However, it is a challenge for emerging regions to spend huge money on theme parks, that bottleneck could be complemented through developing virtual theme parks. On the other hand, the model could be used as a means of enhancing ecotourism in the regions. Especially, reducing the amount of actual visits to the destinations would reduce the environmental effect and possibly contribute towards the development of ecotourism industry. In addition to this, the model could be used as a means to replicate some endangered destinations which are not accessible to human beings.

**References**

Butler, D. (2012). Second Life Machinima enhancing the learning of Law: Lessons from Successful Endeavors, *Australian Journal of Educational Technology*.

Cheong, R. (1995), The Virtual Threat to Travel and Tourism, Tourism Management, vol. 16, no. 6, pp. 417 – 422.

Davis, A., Khazanchi, D, Murphy, J. Zigurs, I and Owens, D. (2009), Avatars, People and Virtual Worlds: Foundations for Research in Metaverses, *Journal of the Association of Information Systems,* vol. 10, no. 2, pp. 90-117.

Dreher, C., Reiners, T., Dreher, N. and Dreher, H. (2011), 3D Virtual Worlds as Collaborative Communities Enhancing Human Endeavors: Innovative Approaches in e-learning, *Proceedings of 3rd IEEE International Conference on Digital Ecosystems and Technologies*, pp. 138-143.

Dreher, C., Rieners, T., Dreher, N. and Dreher, H. (2011), 3D Virtual Worlds Enriching Innovation and Collaboration in Information Systems Research, Development and Commercialization, *Proceedings of 3rd IEEE International Conference on Digital Ecosystems and Technologies*, pp. 168-173.

Eklund, P., Goodall, P., Wray, T, Bunt, B. and Lawson, A. (2009), Designing the Digital Ecosystem of the Virtual Museum of the Pacific, *Proceedings of 3rd IEEE International Conference on Digital Ecosystems and Technologies*, pp. 377-383.

Gregory, S. and Masters, Y. (2012), Real thinking with Virtual Hats: A role playing activity for pre-service teachers in second life, *Australian Journal of Educational Technology*.

Guttentag, D. A. (2010), Virtual Reality: Applications and Implications for Tourism, Tourism Management, vol. 31 (5), pp. 637 651.

Hendaoui, A., Limayem, M., and Thompson, C.W., (2008), 3D Social Virtual Worlds: Research Issues and Challenges, *IEEE Internet Computing*, pp. 88-92.

Henderson, M., Huang, H., Grant, S. and Henderson, L. (2012). The impact of Chinese Language Lessons in a Virtual World on University Students' Self Efficacy Beliefs, *Australian Journal of Educational Technology*.

Kohler, T. Matzler, K and Fuller, J. (2009), Avatar based Innovation: Using Virtual Worlds for real world Innovation, *Technovation*, 29, pp. 395 – 407.

Messinger, P.R., Stroulia, E, Lyons, K. (2008), Virtual Worlds Research: Past, Present & Future, *Journal of Virtual Worlds Research*, vol.1, no.1, pp. 2-18.

Newman, T., W. (2008), Imaginative Travel: Experiential Aspects of User Interactions with Destination Marketing Websites, Doctoral Thesis, Auckland university of Technology, Auckland, New Zealand, Retrieved on 08th September 2009 from http://aut.researchgateway.ac.nz/handle/10292/665?mode=simple.

Pizam, A. (2009). The Global Financial Crisis and Its Impact on the Hospitality Industry. *International Journal of Hospitality Management*, 28, 301.

Steuer, J. (1993), Defining Virtual Reality: Dimensions Determining Telepresence, Journal of Communication, autumn, 1992, pp. 73 – 93.

Stewart, S. and Davis, D. (2012), On the MUVE or in decline: Reflecting on the Sustainability of the Virtual Birth Centre developed in Second Life, *Australian Journal of Educational Technology*.

Tjostheim, Ingvar, Lous, Joachim (2005). A Game Experience in Every Travel Website?

Wegener, M., McIntyre, T.J., McGrath, D., Savage, C.M., Williamson, M., Developing a Virtual Physics World, Australian Journal of Educational Technology, 28 (3), pp. 504-521.

Williams, P. & Hobson, J.S.P., (1995), Virtual Reality and Tourism: Fact or Fantasy?, Tourism Management, vol. 16, no. 6, pp. 423 – 427.

Wood, D. and Willems, J. (2012), Responding to the widening participation agenda through improved access to and within 3D virtual learning environments, *Australian Journal of Educational Technology*.